\documentclass[aps,prd,a4paper,preprintnumbers,superscriptaddress,twocolumn,floatfix,showpacs,nofootinbib]{revtex4}
\usepackage{amssymb}
\usepackage{amsmath}

\usepackage{float,graphicx}

\begin{document}

\title{Detecting a Lorentz-Violating Field in Cosmology}
\author{Baojiu~Li}
\email[Email address: ]{B.Li@damtp.cam.ac.uk}
\affiliation{Department of Applied Mathematics \& Theoretical
Physics, Centre for Mathematical Sciences, University of
Cambridge, Wilberforce Road, Cambridge CB3 0WA, United Kingdom}
\author{David~F.~Mota}
\email[Email address: ]{D.Mota@thphys.uni-heidelberg.de}
\affiliation{Institute of Theoretical Physics, University of Heidelberg, 69120
Heidelberg, Germany}
\author{John~D.~Barrow}
\email[Email address: ]{J.D.Barrow@damtp.cam.ac.uk}
\affiliation{Department of Applied Mathematics \& Theoretical
Physics, Centre for Mathematical Sciences, University of
Cambridge, Wilberforce Road, Cambridge CB3 0WA, United Kingdom}
\date{\today}

\begin{abstract}
We consider cosmology in the Einstein-\AE ther theory (the
generally covariant theory of gravitation coupled to a dynamical
timelike Lorentz-violating vector field) with a linear \AE
-Lagrangian. The $3+1$ spacetime splitting approach is used to
derive covariant and gauge invariant perturbation equations which
are valid for a general class of Lagrangians. Restricting
attention to the parameter space of these theories which is
consistent with local gravity experiments, we show that there are
tracking behaviors for the \AE\ field, both in the background
cosmology and at linear perturbation level. The primordial
power-spectrum of scalar perturbations in this model is shown to
be the same that predicted by standard general relativity.
However, the power-spectrum of tensor perturbation is different
from that in general relativity, but has a smaller amplitude and
so cannot be detected at present. We also study the implications
for late-time cosmology and find that the evolution of photon and
neutrino anisotropic stresses can source the \AE\ field
perturbation during the radiation and matter dominated epochs, and
as a result the CMB and matter power spectra are modified. However
these effects are degenerate with respect to other cosmological
parameters, such as neutrino masses and the bias parameter in the
observed galaxy spectrum.
\end{abstract}

\maketitle

\section{Introduction}

\label{sect:Introduction}

For more than two decades Milgrom's modified Newtonian dynamics (MOND) \cite%
{mond} has been able to explain galaxy rotation curves which are
conventionally considered as an evidence of cold dark matter (CDM) on
galactic scales. MOND modifies Newton's second law of motion to $\mu (|\vec{a%
}|/a_{0})\vec{a}=-\nabla \Phi _{N},$ where $\vec{a}$ and $\Phi
_{N}$ are the acceleration and Newtonian gravitational potential,
respectively; $\mu (x)$ is an effectively free function tending to
unity in the limit $|\vec{a}|\gg a_{0}$, with $a_{0}$ being a new
fundamental constant, which must have a numerical value of
$a_{0}\sim (200\ \mathrm{km/s})^{2}/(10\ \mathrm{kpc})$ in order
to match observations on a galactic scale. This theory looks like
Newton's when accelerations are large but is significantly
different when accelerations are small. On galactic scales,
$|\vec{a}|\ll a_{0}$, so the Newtonian dynamics is modified, but
in a way that can fit spiral galaxy rotation curves. Subsequently,
Bekenstein \cite{TeVeS} built a relativistic theory which has MOND
as a non-relativistic, weak-field limit, thus making the study of
cosmology possible. In addition to the conventional tensor
gravitational field, Bekenstein's theory involves a vector and a
scalar field, and is therefore dubbed TeVeS. Interestingly, it has
been argued that TeVeS could also explain the large-scale
structure formation of the Universe without recurring to CDM
\cite{TeVeSperturbation}, thanks to the presence of the vector
field \cite{TeVeSVector}.

Recently, the authors of Ref.~\cite{TeVeSAether} showed that TeVeS
is equivalent to a vector-tensor theory of gravitation where the
vector field has a non-fixed norm. They also showed that the
correct MONDian limit could be realized with a single vector field
having non-canonical kinetic terms \cite{NoncanonicalAether}.
These results indicate that a vector field in the gravity sector
might be an interesting component of the Universe (indeed the
model in \cite{NoncanonicalAether} and its generalized version we
shall presented below in Eq.~(\ref{eq:ouraction}) could be used to
explain dark energy and dark matter in background cosmology) and
merits more detailed investigations.

The idea of a vector field coupled to gravity has a long worldline
(see for example \cite{AetherReview} for a review and
\cite{others} for further references), but in this work we will
focus on the model described in \cite{AetherModel}, which is the
most well-studied one, and investigate its cosmological
implications. This particular theory is based on a dynamical
vector field coupled to gravitation that picks up a preferred
frame and preserves general covariance. This vector field is
unit-norm, timelike, and violates local Lorentz invariance. It is
called the \AE ther field (or simply \AE -field) and we will refer
to the associated Einstein-\AE ther theory as \AE -theory defined
by the \AE -Lagrangian. The \AE -Lagrangian considered in
\cite{AetherModel} is a special case of our general model
introduced in Eq.~(\ref{eq:ouraction}) below for a unit-norm
vector field that includes terms up to second order in
derivatives, and it has been extensively studied in various
contexts \cite{AetherWF, AetherWF2, AetherCompactStars,
AetherGravitationalWaves, AetherBHs}.

In Refs.~\cite{AetherBackground, AetherPerturbation}, the background
cosmology and primordial power spectra of perturbations from inflation of a
slight different model were also considered. Here, we investigate these for
the model presented in Ref.~\cite{AetherModel}, and also study the evolution
of linear perturbation to the \AE -field during the radiation and
matter-dominated epochs. As we will show below, if we restrict the parameter
space of the underlying theory so as to satisfy the local experimental
gravity constraints, \cite{AetherWF, AetherGravitationalWaves}, this
perturbation becomes sourceless and decays during the epoch of inflation and
late-matter domination. However, it is sourced by the evolutions of the photon and
neutrino anisotropic stresses during the radiation era and early matter era,
which have some imprints on the cosmological observables.

Our presentation is organized as follows. In
Sec.~\ref{sect:Equations}, we briefly introduce the general \AE
-theory and derive perturbation equations for the background
Friedmann-like cosmology in the covariant and gauge invariant
(CGI) formalism (see \cite{Zhao2007a} for a derivation of the
perturbation equations in conformal Newtonian gauge). In
Sec.~\ref{sect:Cosmology} we shall use these equations to discuss
the perturbation dynamics for the cosmological models of
Ref.~\cite{AetherModel}. First, we summarize the existing
constraints on the model in Sec.~\ref{sect:ParameterSpace}; then,
in Sec.~\ref{sect:PerturbationEvolution}, we present the evolution
equations for the perturbation variables and then use them to show
how the primordial spectra of scalar and tensor perturbations in
this theory are unmodified and modified, respectively, on
comparing them with the predictions of general relativity (GR).
The late-time evolution of the \AE -field perturbation and its
effects on cosmological observables are also studied there.
Finally, our discussion and conclusions are presented in
Sec.~\ref{sect:Conclusion}.

Throughout this work our convention is $[\nabla_{a},
\nabla_{b}]u^{c} = R^{\ \ c}_{ab\ d}u^{d}, R_{ab} = R_{acb}^{\ \ \
c}$, where $R_{abcd}, R_{ab}$ are respectively the Riemann tensor
and Ricci tensor; the metric signature is $(+,-,-,-)$ and the
universe is assumed to be spatially flat, filled with photons,
baryons, CDM, 3 species of neutrinos and a cosmological constant.

\section{Field Equations of Einstein-\AE ther Theory}

\label{sect:Equations}

In this section we briefly introduce the general form of the
\text{\AE} -theory and derive the CGI perturbation equations which
we will use to analyze the cosmological effects of the \text{\AE}
-field. The equations presented here are for general Lagrangians.
Later, we shall focus on a specific class of such theory
characterized by a linear Lagrangian (with $f(K)=-K$, see below).

\subsection{The General Einstein-\text{\AE} ther Theory}

The model we consider here is a slight generalization of that presented in
\cite{NoncanonicalAether}. It is characterized by a general gravitational
action of the form
\begin{eqnarray} \label{eq:ouraction}
S &=& \frac{1}{16\pi G_{\mathrm{N}}}\int d^{4}x\sqrt{-g}\left[R + \mathcal{L}_{\text{\AE }} + \lambda(\text{\AE} ^{a}\text{\AE}_{a}-1)\right]
\end{eqnarray}
in which $G_{\mathrm{N}}$ is the \emph{bare} Newtonian gravitation constant,
$\lambda $ is a Lagrange multiplier ensuring that \text{\AE} -field has a unit
norm, and $\mathcal{L}_{\text{\AE} }$ is the \text{\AE} -Lagrangian expressed by
\begin{eqnarray}
\mathcal{L}_{\text{\AE} } &=& f(K^{ab}_{\ \
cd}\nabla_{a}\text{\AE} ^{c}\nabla_{b}\text{\AE}^{d})
\end{eqnarray}
with
\begin{eqnarray}  \label{eq:Kabcd}
K^{ab}_{\ \ cd} &=& c_{1}g^{ab}g_{cd} +
c_{2}\delta^{a}_{c}\delta^{b}_{d} +
c_{3}\delta^{a}_{d}\delta^{b}_{c} +
c_{4}\text{\AE}^{a}\text{\AE}^{b}g_{cd}
\end{eqnarray}
and $f(\cdots)$ being an arbitrary analytic function of its
arguments. As long as the norm of $\text{\AE} _{a}$ is fixed
($\text{\AE} ^{a}\text{\AE} _{a}=1$), the form of $K_{\ \
cd}^{ab}$ is the most general possibility we can have for our
vector-field Lagrangian. Notice that Eq.~(\ref{eq:Kabcd}) here
differs from that given in \cite{NoncanonicalAether} by the
$c_{4}$ term and we shall refer to our model and that of
\cite{NoncanonicalAether} as the GEA (Generalized Einstein \AE
ther) model to distinguish from the one considered in
\cite{AetherModel} (for which we instead called EA (Einstein \AE
ther)). The matter Lagrangian is taken to be the same as in
standard $\Lambda \mathrm{CDM}$ model.

We treat the \text{\AE} -field $\text{\AE}^{a}$ and inverse metric
$g^{ab}$ as the dynamical degrees of freedom and vary the action
with respect to them to obtain the field equations. The former
gives the \text{\AE} -field equation of motion (EOM):
\begin{eqnarray}
\nabla_{b}\left(FJ^{b}_{\ a}\right) -
c_{4}F\text{\AE}^{b}\nabla_{b}\text{\AE}^{c}\nabla_{a}\text{\AE}_{c}
&=& \lambda\text{\AE}_{a}
\end{eqnarray}
where we have defined $K\equiv K_{\ \ cd}^{ab}\nabla
_{a}\text{\AE} ^{c}\nabla _{b}\text{\AE} ^{d}$, $F\equiv
F(K)\equiv \partial f(K)/\partial K$ and $J_{\ c}^{a}\equiv K_{\ \
cd}^{ab}\nabla _{b}\text{\AE} ^{d}$. The variation with respect to
the metric leads to a modified Einstein equation. One could retain
the form of Einstein equations in standard GR by treating the
vector field as a new contribution (denoted by $T_{ab}^{\text{\AE}
}$) to the total energy-momentum tensor in the universe, in
addition to that of the conventional fluid matter which is denoted
by $T_{ab}^{f}$. Then, according to the definition
\begin{eqnarray}
16\pi G_{\mathrm{N}}T^{\text{\AE}}_{ab} &\equiv&
\frac{-2}{\sqrt{-g}}\frac{\delta(\sqrt{-g}\mathcal{L}^{\prime}_{\text{\AE}})}{\delta
g^{ab}}\nonumber
\end{eqnarray}
in which $\mathcal{L}^{\prime }_{\text{\AE} } =
\mathcal{L}_{\text{\AE} } +
\lambda(\text{\AE}^{a}\text{\AE}_{a}-1)$, we have
\begin{widetext}
\begin{eqnarray}
8\pi G_{\mathrm{N}}T^{\text{\AE}}_{ab}&=&
-\nabla_{c}\left\{F\left[\text{\AE}^{c}J_{(ab)} + \text{\AE}_{(a}J^{c}_{\ b)} -
\text{\AE}_{(a}J_{b)}^{\ c}\right]\right\} + \frac{1}{2}g_{ab}f +
\left[\text{\AE}_{d}\nabla_{c}(FJ^{cd}) -
c_{4}F(\text{\AE}^{d}\nabla_{d}\text{\AE}_{c})(\text{\AE}^{e}\nabla_{e}\text{\AE}^{c})\right]\text{\AE}_{a}\text{\AE}_{b}\nonumber\\
&& - F\left[c_{1}\nabla_{a}\text{\AE}^{c}\nabla_{b}\text{\AE}_{c} -
c_{1}\nabla^{c}\text{\AE}_{a}\nabla_{c}\text{\AE}_{b} -
c_{4}(\text{\AE}^{c}\nabla_{c}\text{\AE}_{a})(\text{\AE}^{d}\nabla_{d}\text{\AE}_{b})\right].
\end{eqnarray}
\end{widetext}

We also note that by varying the action with respect to the
Lagrangian multiplier $\lambda$ we simply get the normalization
relation of the \text{\AE} -field, $\text{\AE}^{a}\text{\AE}
_{a}=1$, as mentioned above.

\subsection{The Perturbation Equations in General Relativity}

The CGI perturbation equations in general \text{\AE} -theories are derived in this
section using the method of $3+1$ decomposition \cite{GR3+1} (see \cite%
{MG3+1} for applications of this method in modified-gravity models). First,
we briefly review the main ingredients of $3+1$ decomposition and their
application to standard general relativity \cite{GR3+1} for ease of later
reference.

The main idea of $3+1$ decomposition is to make spacetime splits of physical
quantities with respect to the 4-velocity $u^{a}$ of an observer. The
projection tensor $h_{ab}$ is defined as $h_{ab}=g_{ab}-u_{a}u_{b}$ and can
be used to obtain covariant tensors perpendicular to $u$. For example, the
covariant spatial derivative $\hat{\nabla}$ of a tensor field $T_{d\cdot
\cdot \cdot e}^{b\cdot \cdot \cdot c}$ is defined as
\begin{eqnarray}  \label{eq:AEEOM}
\hat{\nabla}^{a}T_{d\cdot \cdot \cdot e}^{b\cdot \cdot \cdot c}\equiv
h_{i}^{a}h_{j}^{b}\cdot \cdot \cdot \ h_{k}^{c}h_{d}^{r}\cdot \cdot \cdot \
h_{e}^{s}\nabla ^{i}T_{r\cdot \cdot \cdot s}^{j\cdot \cdot \cdot k}.
\end{eqnarray}%
The energy-momentum tensor and covariant derivative of the 4-velocity are
decomposed respectively as
\begin{eqnarray}  \label{eq:AEEMT}
T_{ab} &=&\pi _{ab}+2q_{(a}u_{b)}+\rho u_{a}u_{b}-ph_{ab}, \\
\nabla _{a}u_{b} &=&\sigma _{ab}+\varpi _{ab}+\frac{1}{3}\theta
h_{ab}+u_{a}A_{b}.
\end{eqnarray}%
In the above, $\pi _{ab}$ is the projected symmetric trace-free (PSTF)
anisotropic stress, $q_{a}$ the heat flux vector, $p$ the isotropic
pressure, $\sigma _{ab}$ the PSTF shear tensor, $\varpi _{ab}=\hat{\nabla}%
_{[a}u_{b]}$ the vorticity, $\theta =\nabla ^{c}u_{c}\equiv 3\dot{a}/a$ ($a$
is the mean expansion scale factor) the expansion scalar, and $A_{b}=\dot{u}%
_{b}$ the acceleration; the overdot denotes time derivative expressed as $%
\dot{\phi}=u^{a}\nabla _{a}\phi $, brackets mean antisymmetrisation, and
parentheses symmetrization. The 4-velocity normalization is chosen to be $%
u^{a}u_{a}=1$. The quantities $\pi_{ab},q_{a},\rho ,p$ are referred to as
\emph{dynamical} quantities and $\sigma _{ab},\varpi _{ab},\theta ,A_{a}$ as
\emph{kinematical} quantities. Note that the dynamical quantities can be
obtained from the energy-momentum tensor $T_{ab}$ through the relations
\begin{eqnarray}  \label{eq:DefDynamicalQuantity}
\rho &=&T_{ab}u^{a}u^{b},  \nonumber \\
p &=&-\frac{1}{3}h^{ab}T_{ab},  \nonumber \\
q_{a} &=&h_{a}^{d}u^{c}T_{cd},  \nonumber \\
\pi _{ab} &=&h_{a}^{c}h_{b}^{d}T_{cd}+ph_{ab}.
\end{eqnarray}

Decomposing the Riemann tensor and making use the Einstein equations, we
obtain, after linearization, five constraint equations \cite{GR3+1}:
\begin{eqnarray}
\label{eq:ConstraintVarpi} 0 &=&\hat{\nabla}^{c}(\varepsilon _{\ \ cd}^{ab}u^{d}\varpi _{ab}); \\
\label{eq:Constraintq} \kappa q_{a} &=&
-\frac{2\hat{\nabla}_{a}\theta}{3} +
\hat{\nabla}^{b}\sigma_{ab}+\hat{\nabla}^{b}\varpi _{ab};\ \ \  \\
\mathcal{B}_{ab} &=&\left[ \hat{\nabla}^{c}\sigma
_{d(a}+\hat{\nabla}^{c}\varpi _{d(a}\right] \varepsilon _{b)ec}^{\ \ \ \ d}u^{e}; \\
\label{eq:Constraintphi} \hat{\nabla}^{b}\mathcal{E}_{ab} &=&
\frac{1}{2}\kappa \left[\hat{\nabla}^{b}\pi_{ab}+\frac{2}{3}\theta
q_{a}+\frac{2}{3}\hat{\nabla}_{a}\rho \right];\\
\hat{\nabla}^{b}\mathcal{B}_{ab} &=&\frac{1}{2}\kappa
\left[\hat{\nabla}_{c}q_{d}+(\rho +p)\varpi _{cd}\right]
\varepsilon _{ab}^{\ \ cd}u^{b},
\end{eqnarray}
and five propagation equations,
\begin{eqnarray}
\label{eq:Raychaudhrui} \dot{\theta}+\frac{1}{3}\theta^{2}
-\hat{\nabla}^{a}A_{a}+\frac{\kappa }{2}(\rho +3p) &=& 0; \\
\label{eq:Propagationsigma} \dot{\sigma}_{ab}+\frac{2}{3}\theta
\sigma _{ab}-\hat{\nabla}_{\langle
a}A_{b\rangle }+\mathcal{E}_{ab}+\frac{1}{2}\kappa \pi _{ab} &=& 0; \\
\dot{\varpi}+\frac{2}{3}\theta \varpi -\hat{\nabla}_{[a}A_{b]} &=& 0; \\
\label{eq:Propagationphi} \frac{1}{2}\kappa \left[\dot{\pi}_{ab} +
\frac{1}{3}\theta\pi_{ab}\right] - \frac{1}{2}\kappa \left[(\rho
+p)\sigma_{ab}+\hat{\nabla}_{\langle
a}q_{b\rangle}\right]  \nonumber \\
-\left[ \dot{\mathcal{E}}_{ab}+\theta \mathcal{E}_{ab}-\hat{\nabla}^{c}%
\mathcal{B}_{d(a}\varepsilon _{b)ec}^{\ \ \ \ d}u^{e}\right] &=&
0;\ \ \ \
\\
\dot{\mathcal{B}}_{ab}+\theta \mathcal{B}_{ab}+\hat{\nabla}^{c}\mathcal{E}%
_{d(a}\varepsilon _{b)ec}^{\ \ \ \ d}u^{e}  \nonumber \\
+\frac{\kappa }{2}\hat{\nabla}^{c}\mathcal{\pi }_{d(a}\varepsilon
_{b)ec}^{\ \ \ \ d}u^{e} &=& 0.
\end{eqnarray}
Here, $\varepsilon _{abcd}$ is the covariant permutation tensor, $\mathcal{E}%
_{ab}$ and $\mathcal{B}_{ab}$ are respectively the electric and magnetic
parts of the Weyl tensor $\mathcal{W}_{abcd}$, defined by $\mathcal{E}%
_{ab}=u^{c}u^{d}\mathcal{W}_{acbd}$ and $\mathcal{B}_{ab}=-\frac{1}{2}%
u^{c}u^{d}\varepsilon _{ac}^{\ \ ef}\mathcal{W}_{efbd}$. The angle bracket
means taking the trace-free part of a quantity.

Besides the above equations, it is useful to express the projected Ricci
scalar $\hat{R}$ into the hypersurfaces orthogonal to $u^{a}$ as
\begin{eqnarray}  \label{eq:SpatialRicciCurvature}
\hat{R} &\doteq& 2\kappa\rho - \frac{2}{3}\theta ^{2}.
\end{eqnarray}%
The spatial derivative of the projected Ricci scalar, $\eta _{a}\equiv \frac{%
1}{2}a\hat{\nabla}_{a}\hat{R}$, is then given as
\begin{eqnarray}  \label{eq:Constrainteta}
\eta _{a} &=& \kappa \hat{\nabla}_{a}\rho - \frac{2a}{3}\theta\hat{\nabla}%
_{a}\theta,
\end{eqnarray}%
and its propagation equation by
\begin{eqnarray}  \label{eq:Propagationeta}
\dot{\eta}_{a}+\frac{2\theta }{3}\eta _{a} &=& -\frac{2}{3}\theta a\hat{%
\nabla}_{a}\hat{\nabla}\cdot A - a\kappa\hat{\nabla}_{a}\hat{\nabla}\cdot q.
\end{eqnarray}

Finally, there are the conservation equations for the energy-momentum
tensor:
\begin{eqnarray}  \label{eq:EnergyConservation}
\dot{\rho}+(\rho +p)\theta +\hat{\nabla}^{a}q_{a} &=& 0, \\
\label{eq:HeatfluxEvolution} \dot{q}_{a}+\frac{4}{3}\theta q_{a}+(\rho +p)A_{a}-\hat{\nabla}_{a}p+\hat{%
\nabla}^{b}\pi _{ab} &=& 0.
\end{eqnarray}

As we are considering a spatially-flat universe, the spatial
curvature must vanish on large scales and so $\hat{R}=0$. Thus,
from Eq.~(\ref{eq:SpatialRicciCurvature}), we obtain
\begin{eqnarray}
\frac{1}{3}\theta ^{2}=\kappa\rho.
\end{eqnarray}
This is the Friedmann equation in standard general relativity, and the other
background equations (the Raychaudhuri equation and the energy-conservation
equation) can be obtained by taking the zero-order parts of Eqs.~(\ref{eq:Raychaudhrui}, \ref{eq:EnergyConservation}), yielding:
\begin{eqnarray}
\dot{\theta}+\frac{1}{3}\theta ^{2}+\frac{\kappa }{2}(\rho +3p) &=& 0, \\
\label{eq:BackgroundEnergyConservation} \dot{\rho}+(\rho +p)\theta
&=& 0.
\end{eqnarray}

In what follows, we will only consider scalar perturbation modes, for which
the vorticity $\varpi _{ab}$ and magnetic part of Weyl tensor $\mathcal{B}%
_{ab}$ are at most of second order \cite{GR3+1}, and so will be neglected in
our first-order analysis.

\subsection{The Perturbation Quantities in \text{\AE} -Theory}

In the Einstein-\text{\AE}ther theories where we consider the
\text{\AE} -field as a new
species of matter, the gravitational field equations Eqs.~(\ref%
{eq:ConstraintVarpi} - \ref{eq:BackgroundEnergyConservation}) listed above
preserve their forms, but the dynamical quantities $\rho ,p,q_{a},\pi _{ab}$
appearing there should be replaced by the effective total quantities of the
same type. For simplicity, we shall \emph{always} use variables \emph{without%
} superscripts to denote these \emph{effective total} quantities, while for
those of a specified matter species we shall add corresponding superscripts (%
\emph{e.g.}, $\rho ^{\text{\AE} }$ denotes the energy density of
the \text{\AE} -field $\cdots$).

The vector \text{\AE} -field, $\text{\AE} _{a}$, requires further discussion. As we
mentioned above, it has the normalization relation $\text{\AE} ^{a}\text{\AE} %
_{a}=u^{a}u_{a}=1$. In the background Friedmann-Robertson-Walker (FRW)
universe the requirements of homogeneity and isotropy require that $\text{\AE} _{a}$
is just equal to $u_{a}$, which is unambiguously chosen as the 4-velocity of
the fundamental observers. But in a perturbed, almost-FRW, universe this is
no longer true and we can write $\text{\AE} _{a}=u_{a}+\text{\ae} _{a},\text{\AE} ^{a}=u^{a}+\text{\ae} %
^{a}$ where $\text{\ae} _{a}$ is another (first-order) vector field that vanishes
in a FRW Universe: we call it the perturbation of the \text{\AE} -field.
Furthermore, the relation $\text{\AE} ^{a}\text{\AE} _{a}=u^{a}u_{a}=1$ implies that $u^{a}%
\text{\ae} _{a}=0$, \emph{i.e.}, $\text{\ae} _{a}$ is a \emph{spatial} vector field which
is perpendicular to $u_{a}$, up to first order in perturbation. This fact is
used extensively in deriving the perturbation equations (\emph{e.g.}, $%
\nabla ^{a}\text{\ae} _{a}=\hat{\nabla}^{a}\text{\ae} _{a}$ \emph{etc}.).

With these preliminaries at hand, and after some lengthy manipulations, the
\text{\AE} -field EOM Eq.~(\ref{eq:AEEOM}) can be written as (up to first order)
\begin{widetext}
\begin{eqnarray} \label{eq:aeEOM}
c_{14}\left[F\ddot{\text{\ae}}_{a}+\left(\dot{F}
+F\theta\right)\dot{\text{\ae}}_{a}\right]+ c_{14}
\left[F\dot{A}_{a}+\dot{F}A_{a}+\frac{2}{3}F\theta A_{a}\right] -
\left[\frac{1}{3}(\alpha-c_{14})(\dot{F}\theta +F\dot{\theta}) -
\frac{2}{9}c_{14}F\theta^{2}\right]\text{\ae}_{a}\nonumber\\
+ \frac{1}{3}\alpha\hat{\nabla}_{a}(F\theta) + \frac{1}{3}\alpha
F\hat{\nabla}_{a}\hat{\nabla}^{b}\text{\ae}_{b} +
c_{13}F\hat{\nabla}^{b}\left(\sigma_{ab}+\hat{\nabla}_{\langle
a}\text{\ae}_{b\rangle}\right) &=& 0,
\end{eqnarray}
\end{widetext}
and from the definitions Eq.~(\ref{eq:DefDynamicalQuantity}) the \text{\AE} -field
energy density, isotropic pressure, heat-flux vector and anisotropic stress
can be identified from Eq.~(\ref{eq:AEEMT}) (again up to first order) as
\begin{widetext}
\begin{eqnarray}
\label{eq:AErho} \kappa\rho^{\text{\AE}} &=& \frac{1}{2}f -
\frac{1}{3}F\alpha(\theta^{2}+2\theta\hat{\nabla}^{a}\text{\ae}_{a}) +
c_{14}F\hat{\nabla}^{a}\left(A_{a} + \dot{\text{\ae}}_{a} +
\frac{1}{3}\theta\text{\ae}_{a}\right),\\
\label{eq:AEp} \kappa p^{\text{\AE}} &=& -\frac{1}{2}f +
\frac{\alpha}{3}\dot{F}(\theta + \hat{\nabla}^{a}\text{\ae}_{a}) +
\frac{\alpha}{3}F \left[\dot{\theta} + \theta^{2} +
(\hat{\nabla}^{a}\text{\ae}_{a})^{\cdot} +
2\theta\hat{\nabla}^{a}\text{\ae}_{a}\right],\\
\label{eq:AEq} \kappa q^{\text{\AE}}_{a} &=& -c_{14}
\left[F\dot{A}_{a}+\dot{F}A_{a}+\frac{2}{3}F\theta A_{a}\right]
-c_{14}\left[F\ddot{\text{\ae}}_{a}+\left(\dot{F}
+F\theta\right)\dot{\text{\ae}}_{a}\right] +
\left[\frac{1}{3}(\alpha-c_{14})(\dot{F}\theta
+F\dot{\theta}) - \frac{2}{9}c_{14}F\theta^{2}\right]\text{\ae}_{a}\\
\label{eq:AEq2} &=& \frac{1}{3}\alpha\hat{\nabla}_{a}(F\theta) +
\frac{1}{3}\alpha F\hat{\nabla}_{a}\hat{\nabla}^{b}\text{\ae}_{b} +
c_{13}F\hat{\nabla}^{b}\left(\sigma_{ab}+\hat{\nabla}_{\langle
a}\text{\ae}_{b\rangle}\right),\\
\label{eq:AEpi} \kappa\pi^{\text{\AE}}_{ab} &=&
-c_{13}(\dot{F}+F\theta)\left[\sigma_{ab} +\hat{\nabla}_{\langle
a}\text{\ae}_{b\rangle}\right] - c_{13}F\left[\dot{\sigma}_{ab}
+(\hat{\nabla}_{\langle a}\text{\ae}_{b\rangle})^{\cdot}\right],
\end{eqnarray}
\end{widetext}
where, in Eq.~(\ref{eq:AEq2}), we have used Eq.~(\ref{eq:aeEOM}). Here, we
have defined the new parameters $\alpha \equiv c_{1}+3c_{2}+c_{3}$, $%
c_{13}\equiv c_{1}+c_{3}$ and $c_{14}\equiv c_{1}+c_{4} $; $\kappa $ is
given by $\kappa =8\pi G_{\mathrm{N}}$. Including these \text{\AE} -contributions
to Eqs.~(\ref{eq:ConstraintVarpi} - \ref{eq:BackgroundEnergyConservation})
one obtains the modified gravitational field equations for the general \text{\AE} %
-theory. It is also easy to check that the above results satisfy
(separately) the conservation of the \text{\AE} -field's
energy-momentum tensor, Eqs.~(\ref{eq:EnergyConservation},
\ref{eq:HeatfluxEvolution}). Note that Eqs.~(\ref{eq:AErho} -
\ref{eq:AEpi}) are the general expressions of energy density,
pressure, heat flux and anisotropic stress in the $3+1$
decomposition which include both zeroth order (background) and
first order terms; to calculate the actually density contrast
\emph{etc.} (see Eqs.~(\ref{eq:AErho_2} - \ref{eq:AEpi_2}) below)
one needs to take the covariant spatial derivatives of these
equations \cite{GR3+1}.

\section{A Specific Model: The linear Lagrangian}

\label{sect:Cosmology}

In above we presented the field equations for general \text{\AE} -theories, but in
what follows we shall only analyze the cosmology of a specific edition of
the theory which is defined by choice of a linear Lagrangian:
\begin{eqnarray}  \label{eq:AELagrangian}
f(K) &=& -K
\end{eqnarray}
(note that the minus sign is because of our sign convention). This
model is by far the most well known, in the sense that it has been
investigated in the contexts of static weak-field limit
\cite{AetherWF} (see also \cite{AetherWF2} for the weak-field
limit of the \AE -model considered in \cite{NoncanonicalAether}),
background
cosmology \cite{AetherBackground}, the radiation and propagation of the \text{\AE} %
-gravitational waves \cite{AetherPerturbation, AetherGravitationalWaves},
compact stars \cite{AetherCompactStars}, and black holes \cite{AetherBHs}.
Some of these studies have imposed stringent constraints on the viable
parameter-space of $c_{i}$s. In view of these restrictions we will confine
our study to this constrained subset of possible theories. Note that the
perturbation dynamics of the \text{\AE} -model has also been analyzed in \cite%
{AetherPerturbation} in the absence of the $c_{4}$ term. Here, we shall
include this term and perform a similar analysis but in slightly different
manner and in more detail; we will also discuss on some detailed features
which lead to our conclusions being different. The late-time evolution of
the \text{\AE} -perturbation is also investigated. In particular, we shall find
that within the locally-constrained parameter space the \text{\AE} -model will
leave slightly different signatures on the perturbation evolutions from
those left by the standard $\Lambda \mathrm{CDM}$ paradigm in general
relativity, and so cosmological data on cosmic microwave background (CMB)
and matter power spectra might place some constraint on the parameter space.

\subsection{The Constrained Parameter Space}

\label{sect:ParameterSpace}

In this subsection we briefly summarize the constraints on the \text{\AE} -model
described in Eq.~(\ref{eq:AELagrangian}). It has been well known that in the
weak-field, slow-motion, limit and in the background cosmology the \text{\AE} %
-model displays tracking behavior. For the former environment, it can be
shown that in the presence of the \text{\AE} -field the observed \emph{effective}
gravitational constant $G_{0}$ is a rescaling of the bare one $G_{\mathrm{N}%
} $ \cite{AetherBackground, AetherWF} by (note that here our $c_{i}$s have
different signs from those in \cite{AetherWF})
\begin{eqnarray}
G_{0} &=& \frac{1}{1+\frac{1}{2}c_{14}}G_{\mathrm{N}},
\end{eqnarray}
and in the latter environment the observed gravitational constant
$G_{\infty}$ is also a rescaling of $G_{\mathrm{N}},$ but with a
different factor \cite{AetherBackground}
\begin{eqnarray}
G_{\infty} &=& \frac{1}{1-\frac{1}{2}\alpha}G_{\mathrm{N}}.
\end{eqnarray}
Correspondingly, we define $\kappa _{0}=8\pi G_{0}$ and $\kappa
_{\infty }=8\pi G_{\infty }$ to be used below. Note that the
rescaled $G_{\infty }$ is generally not equal to $G_{\mathrm{N}}$
and as such the background cosmic expansion rate will be different
from that in standard GR. However, we note that the numerical
value of gravitational constant we find in the textbooks and use
in the numerical calculations is \emph{not} $G_{\mathrm{N}}$ but
rather the locally-measured $G_{0}$. It is possible to obtain
limits on non-local values of $G$ by considering the primordial
nucleosynthesis of light elements (see for example
Ref.~\cite{bs}). If $c_{14}=-\alpha$ then we have $G_{\infty
}=G_{0}$, which indicates that the background cosmological
dynamics will be exactly the same as \emph{assuming} standard GR
\cite{AetherBackground, Comment}; otherwise we can use constraints
from primordial nucleosynthesis on the value of gravitational
constant to show that $|G_{\infty }/G_{0}-1|\lesssim
\mathcal{O}(0.1)$ \cite{AetherWF}. However, note that the
particle-horizon size at the epoch of neutron-proton freeze-out
($t\sim 1\ \mathrm{s}$), which is most sensitive to variations in
the value of $G$, is only $\sim 10^{10}\ \mathrm{cm}$ and this
causally linked region expands in size by about a factor of
$10^{10}$ by the present-day to a size $\sim 10^{20}\ \mathrm{cm}
\sim 32\ \mathrm{pc}$ which is a sub-galactic scale but there has
then been local gravitational collapse by a factor of $10^{2}$.
Such collapse may however also affect the local value of the
gravitational constant \cite{tim}, but we will not consider this
in the present work.

There are also constraints from the observations of parameterized
post-Newtonian (PPN) parameters \cite{PPNGR}. It is shown in \cite{AetherWF}
that for all the PPN parameters to coincide with those in GR (otherwise the
parameters may need to be fine-tuned) one reduces the full four-parameter
space of the model to a two-parameter subspace characterized by
\begin{eqnarray}  \label{eq:ParameterSpace}
c_{2} &=& \frac{-2c_{1}^{2}-c_{1}c_{3}+c_{3}^{2}}{3c_{1}},\nonumber \\
c_{4} &=& -\frac{c_{3}^{2}}{c_{1}}.
\end{eqnarray}
In addition, \text{\AE} -theories contain five gravitational and \text{\AE} -wave modes,
which include the two usual spin-2 gravitational waves, and three additional
modes: two spin-1 transverse \text{\AE} -gravity waves and one spin-0 longitudinal
\text{\AE} -gravity wave. The speeds of the three additional modes are generally
not equal to $1$. It has been shown in Ref.~\cite{AetherGravitationalWaves}
that if these speeds are less than 1 then the high-energy particles will
produce $\mathrm{\breve{C}}$erenkov radiation when passing through vacuum,
which imposes stringent constraints on the model. However, as suggested in
\cite{AetherWF}, these constraints do not apply if these modes propagate
superluminally. The requirement that the additional \text{\AE} -gravity waves do
not propagate subluminally further limits the parameter space to
\begin{eqnarray}  \label{eq:ParameterSpace2}
-1 < c_{13} < 0,\qquad \frac{c_{13}}{3(1+c_{13})} < c_{1}-c_{3} < 0.
\end{eqnarray}

Finally, when the above constraints Eq.~(\ref{eq:ParameterSpace2})
are satisfied, the positive energy requirement and the stability
of additional wave modes \cite{AetherPerturbation} also hold. In
addition, we will have $c_{14}=-\alpha $ so that the Big Bang
nucleosynthesis constraint does not apply. Thus, we can see that,
even after using all the current constraints, there is still a
large parameter space remaining for the model. In what follows we
shall ask whether linear-order cosmological observations such as
the CMB and the form of the matter power spectrum could reduce
this parameter space further, and as we will show, the answer is
positive. However, the modifications are small and depend weakly
on the model parameters, which mean that the data on CMB and
matter power spectra cannot give very stringent limits on the
parameter space.

\subsection{Linearly Perturbed Equations}

\label{sect:PerturbationEvolution}

In this subsection we consider the perturbation evolutions of our \text{\AE} %
-model. For generality, we will derive the equations for arbitrary
choices of parameters and only later confine ourselves to the
parameter space described in Eq.~(\ref{eq:ParameterSpace2}).
Besides, since the presence of the \text{\AE} -field in general
will modify the cosmology at all times, we will also investigate
its effects during the inflationary era as in
\cite{AetherPerturbation}; after that we will turn to its effects
on late time cosmology.


Following \cite{GR3+1}, we shall make the following harmonic expansions of
our perturbation variables
\begin{eqnarray}  \label{eq:HarmonicExpansion}
\hat{\nabla}_{a}\rho = \sum_{k}\frac{k}{a}\mathcal{X}Q_{a}^{k},\qquad \hat{%
\nabla}_{a}p = \sum_{k}\frac{k}{a}\mathcal{X}^{p}Q_{a}^{k}  \nonumber \\
q_{a} = \sum_{k}qQ_{a}^{k},\qquad \pi_{ab} = \sum_{k}\Pi Q_{ab}^{k},\qquad
\nonumber \\
\hat{\nabla}_{a}\theta = \sum_{k}\frac{k^{2}}{a^{2}}\mathcal{Z}Q_{a}^{k}, \qquad \sigma_{ab} = \sum_{k}\frac{k}{a}\sigma Q_{ab}^{k}  \nonumber \\
\hat{\nabla}_{a}a = \sum_{k}khQ_{a}^{k},\qquad A_{a} = \sum_{k}\frac{k}{a}%
AQ^{k}_{a}  \nonumber \\
\text{\ae} _{a} = \sum_{k}\text{\ae} Q^{k}_{a},\qquad \eta_{a} = \sum_{k}\frac{k^{3}}{a^{2}%
}\eta Q_{a}^{k}  \nonumber \\
\mathcal{E}_{ab} = -\sum_{k}\frac{k^{2}}{a^{2}}\phi Q_{ab}^{k}
\end{eqnarray}
in which $Q^{k}$ is the eigenfunction of the comoving spatial Laplacian $%
a^{2}\hat{\nabla}^{2}$ satisfying
\begin{eqnarray}
\hat{\nabla}^{2}Q^{k} &=& \frac{k^{2}}{a^{2}}Q^{k}  \nonumber
\end{eqnarray}
and $Q_{a}^{k},Q_{ab}^{k}$ are given by $Q_{a}^{k}=\frac{a}{k}\hat{\nabla}%
_{a}Q^{k},Q_{ab}^{k}=\frac{a}{k}\hat{\nabla}_{\langle a}Q_{b\rangle }^{k}$.
Note that $\text{\ae} $ is dimensionless.

In terms of the above harmonic expansion coefficients, Eqs.~(\ref%
{eq:Constraintq}, \ref{eq:Constraintphi}, \ref{eq:Propagationsigma}, \ref%
{eq:Propagationphi}, \ref{eq:Constrainteta}, \ref{eq:Propagationeta}) can be
rewritten as \cite{GR3+1}
\begin{eqnarray}
\label{eq:Constraintq2} \frac{2}{3}k^{2}(\sigma - \mathcal{Z}) &=&
\kappa
qa^{2},\\
\label{eq:Constraintphi2} k^{3}\phi &=& -\frac{1}{2}\kappa
a^{2}\left[k(\Pi
+\mathcal{X})+3\mathcal{H}q\right],\\
\label{eq:Propagationsigma2} k(\sigma' + \mathcal{H}\sigma) &=&
k^{2}(\phi+A)
-\frac{1}{2}\kappa a^{2}\Pi,\\
\label{eq:Propagationphi2} k^{2}(\phi'+\mathcal{H}\phi) &=&
\frac{1}{2}\kappa a^{2}\left[k(\rho+p)\sigma+kq-\Pi'
-\mathcal{H}\Pi\right],\ \ \ \ \\
\label{eq:Constrainteta2} k^{2}\eta &=& \kappa\mathcal{X}a^{2} -
2k\mathcal{H}\mathcal{Z},\\
\label{eq:Propagationeta2} k\eta' &=& -\kappa qa^{2} -
2k\mathcal{H}A
\end{eqnarray}
where $\mathcal{H}=a^{\prime }/a=\frac{1}{3}a\theta $ and a prime
denotes the derivative with respect to the conformal time $\tau$
($ad\tau =dt$). Also, Eq.~(\ref{eq:HeatfluxEvolution}) and the
spatial derivative of Eq.~(\ref{eq:EnergyConservation}) become
\begin{eqnarray}
\label{eq:HeatfluxEvolution2} q' + 4\mathcal{H}q + (\rho+p)kA -
k\mathcal{X}^{p} +
\frac{2}{3}k\Pi &=& 0,\\
\label{eq:EnergyConservation2} \mathcal{X}' + 3h'(\rho+p) +
3\mathcal{H}(\mathcal{X}+\mathcal{X}^{p}) + kq &=& 0
\end{eqnarray}
Recall that we shall always neglect the superscript
$^{\mathrm{tot}}$ for the total dynamical quantities and add
appropriate superscripts for individual matter species. Note that
\begin{eqnarray}
h^{\prime }&=& \frac{1}{3}k\mathcal{Z} - \mathcal{H}A
\end{eqnarray}
and that the $\kappa $ appearing above is the bare (not necessarily the
measured) one. Furthermore, for convenience we define the frame-independent
(FI) variables \cite{GR3+1}
\begin{eqnarray}
\tilde{q} &=& q + (\rho+p)\sigma, \\
\tilde{\text{\ae} } &=& \text{\ae} + \sigma,
\end{eqnarray}
$\tilde{\text{\ae} }$ is FI according to Eq.~(\ref{eq:AEpi})
because we know that the anisotropic pressure tensor $\pi_{ab}$ is
frame invariant. Hence, it follows from
Eq.~(\ref{eq:Constraintphi}), that $\tilde{q}$ is also FI up to
first order in perturbation. In the zero-shear frame (the
Newtonian gauge), we have simply $\tilde{q}=q$ and
$\tilde{\text{\ae} }=\text{\ae}$.

Before presenting the evolution equations, we first write the dynamical
quantities of the \text{\AE} -field in terms of the harmonic coefficients
introduced above:
\begin{eqnarray} \label{eq:AErho_2}
\kappa\mathcal{X}^{\text{\AE} }a^{2} &=& \alpha
k\mathcal{H}(\mathcal{Z}+\text{\ae} )
\nonumber \\
&& - c_{14}\left[k^{2}A + k(\text{\ae} ^{\prime }+\mathcal{H}\text{\ae} )\right],\\
\label{eq:AEp_2} \kappa\mathcal{X}^{p, \text{\AE} }a^{2} &=& -
\frac{\alpha}{3}k\left[(\mathcal{Z}+
\text{\ae})^{\prime }+ 2\mathcal{H}(\mathcal{Z}+\text{\ae} )\right]\nonumber\\
&& - \alpha(\mathcal{H}^{\prime }-\mathcal{H}^{2})A, \\
\label{eq:AEq_2} \kappa q^{\text{\AE} }a^{2} &=&
-\frac{1}{3}\alpha
k^{2}(\mathcal{Z}+\text{\ae} ) - \frac{2}{3}c_{13}k^{2}\tilde{\text{\ae}},\\
\label{eq:AEpi_2} \kappa\Pi^{\text{\AE} }a^{2} &=&
c_{13}k(\tilde{\text{\ae}}^{\prime }+
2\mathcal{H}\tilde{\text{\ae} }).
\end{eqnarray}
In these expressions we have used both $\text{\ae} $ and
$\tilde{\text{\ae}}$ because not all these quantities are FI.
However it can be shown that the two quantities
\begin{eqnarray}  \label{eq:FIAdiabatic}
&& \kappa a^{2}\left(\mathcal{X}^{p, \text{\AE} } - \frac{p^{\prime }}{\rho^{\prime
}}\mathcal{X}^{\text{\AE} }\right)  \nonumber \\
&=& \alpha\left[\frac{1}{2}\kappa a^{2}\left(\mathcal{X}^{p} -
\frac{p^{\prime }}{\rho^{\prime }}\mathcal{X}\right) - k^{2}\frac{p^{\prime }}{%
\rho^{\prime }}\Phi + \frac{\Xi^{\prime }}{3\mathcal{H}\Xi} \kappa a^{2}\Pi%
\right]  \nonumber \\
&& - \left(\frac{1}{3} + \frac{\Xi^{\prime }}{3\mathcal{H}\Xi}\right)c_{14}
\left(k\tilde{\text{\ae} }^{\prime }+ k\mathcal{H}\tilde{\text{\ae} } - k^{2}\Phi\right)
\nonumber \\
&& - \frac{1}{3}\alpha k\tilde{\text{\ae} }^{\prime }- \frac{1}{3}\alpha k\mathcal{H%
}\tilde{\text{\ae} } + \frac{\Xi^{\prime }}{3\mathcal{H}\Xi}\alpha k\mathcal{H}%
\tilde{\text{\ae} }
\end{eqnarray}
and
\begin{eqnarray}  \label{eq:FIAetherq}
\kappa a^{2}\tilde{q}^{\text{\AE} } &\equiv& \kappa a^{2}[(\rho^{\text{\AE} } + p^{\text{\AE} %
})\sigma + q^{\text{\AE} }]  \nonumber \\
&=& \frac{1}{2}\alpha\kappa a^{2}\tilde{q} - \frac{1}{3}\alpha k^{2}\tilde{%
\text{\ae} } - \frac{2}{3}c_{13}k^{2}\tilde{\text{\ae} },
\end{eqnarray}
which will be used in the derivations, are FI, as they are
expressed in terms of FI variables only. Note that in the above we
have defined $\Phi \equiv \phi - \frac{\kappa a^{2}}{2k^{2}}\Pi $
for convenience, where $\Phi $ is the Newtonian gravitational
potential, and $$\Xi \equiv \frac{1}{2}\kappa a^{2}(\rho + p) =
\mathcal{H}^{2} - \mathcal{H}^{\prime }, \quad p^{\prime
}/\rho^{\prime }= -\frac{1}{3} - \frac{\Xi'}{3\mathcal{H}\Xi}.$$

We now investigate in detail Eqs.~(\ref{eq:FIAdiabatic},
\ref{eq:FIAetherq}). On large scales, where $k\tau \ll 1$, the
terms involving $k$ (these include the $\Pi$ term in
Eq.~(\ref{eq:FIAdiabatic})) can be safely disregarded, and as a
result we have
\begin{eqnarray}
\kappa\left(\mathcal{X}^{p, \AE} -
\frac{p'^{\AE}}{\rho'^{\AE}}\mathcal{X}^{\AE}\right) &\simeq&
\frac{\frac{1}{2}\alpha}{1-\frac{1}{2}\alpha}\kappa\left(\mathcal{X}^{p,
f} -
\frac{p'^{f}}{\rho'^{f}}\mathcal{X}^{f}\right),\\
\kappa\tilde{q}^{\AE} &\simeq&
\frac{\frac{1}{2}\alpha}{1-\frac{1}{2}\alpha} \kappa\tilde{q}^{f}
\end{eqnarray}
where the superscript $^{f}$ means the fluid matter. We see that
on large scales these attributes of the \text{\AE} -field track
those for other matter
species in the universe. As the combination $\mathcal{X}^{p,f}-\frac{%
p^{\prime }}{\rho^{\prime }}\mathcal{X}^{f}$ determines the type of
perturbations (for example, the perturbation is adiabatic if the combination
is equal to zero), this indicates that the \text{\AE} -field will not alter the
type of the scalar perturbation produced in the inflationary era. Note that
the above tracking behaviors are the same as that in the background
cosmology, \emph{i.e.}, rescaling the gravitational constant by a same
factor.

Now we can proceed to derive the evolution equations for our \text{\AE} -model.
The propagation equation for the \text{\AE} -field Eq.~(\ref{eq:aeEOM}), in terms
of the FI variables, becomes
\begin{eqnarray}  \label{eq:Propagationepsilon}
(c_{14}-c_{13}\alpha)\epsilon^{\prime \prime }- \alpha(1+c_{13})(\mathcal{H}%
^{\prime }-\mathcal{H}^{2})\epsilon  \\ \nonumber
+ \frac{\alpha+2c_{13}}{3}k^{2}\epsilon - (\alpha+c_{14})k(a\Phi)^{\prime }-
\frac{\alpha}{k} a(\kappa\Pi^{f}a^{2})^{\prime }&=& 0
\end{eqnarray}
in which we have changed the variable to $\epsilon \equiv
a\tilde{\text{\ae} }$ for simplicity. Taking the time derivative
of Eq.~(\ref{eq:Propagationphi2}), adding to it
$(4+3\frac{p^{\prime }}{\rho^{\prime }})\mathcal{H}$ times the
same equation, and using Eq.~(\ref{eq:HeatfluxEvolution2}) to
eliminate the $q^{\prime }$ term, we  obtain the following
equation second-order differential equation for $\Phi$:
\begin{widetext}
\begin{eqnarray} \label{eq:PropagationPhi}
&& \Phi'' + \left(2\mathcal{H}-\frac{\Xi'}{\Xi}\right)\Phi' +
\left(2\mathcal{H}'-\frac{\Xi'}{\Xi}\mathcal{H}\right)\Phi +
k^{2}\frac{p'}{\rho'}\Phi\nonumber\\
&=& \frac{1}{2}\kappa a^{2}\left(\mathcal{X}^{p} -
\frac{p'}{\rho'}\mathcal{X}\right) - \frac{1}{k^{2}}\kappa
a^{2}\left[\Pi'' +
\left(5-\frac{\Xi'}{\mathcal{H}\Xi}\right)\mathcal{H}\Pi' +
2\mathcal{H}'\Pi +
\left(6-2\frac{\Xi'}{\mathcal{H}\Xi}\right)\mathcal{H}^{2}\Pi -
\frac{\Xi'}{3\mathcal{H}\Xi}k^{2}\Pi\right].
\end{eqnarray}
\end{widetext}
Eqs.~(\ref{eq:Propagationepsilon}, \ref{eq:PropagationPhi}) are
the evolution equations for the coupled $\epsilon -\Phi $ system
that we are looking for (note that $\Pi =\Pi^{\text{\AE}} +
\Pi^{f}$ where $\Pi^{f}$ is the fluid matter anisotropic stress
and $\Pi^{\AE}$ can also be expressed in terms of $\epsilon $ and
its time derivatives by virtue of Eq.~(\ref{eq:AEpi_2})). They are
not closed if $\mathcal{X}^{p,f}-\frac{p^{\prime f}}{\rho^{\prime
f}}\mathcal{X}^{f}$ and $\Pi^{f}$ are unknown and to know these
quantitie we would need to know the matter content of the
universe.

\subsection{The Primordial Power Spectra}

In the analysis above, we have mentioned that the presence of the \text{\AE} -field
does not affect the form of the scalar perturbation produced during
inflation. But we also need to know whether other features of the
inflationary power spectrum, such as the spectral index and the amplitude,
are modified by the \text{\AE} -field, as compared with the predictions in standard
GR. Here, we will investigated this issue (see \cite{AetherPerturbation} for
a study in the absence of the $c_{4}$ term in the \text{\AE} -Lagrangian) by
considering a single-field model of inflation \cite{InflationBook} in the
presence of the \text{\AE} -field.

During the inflationary epoch a scalar inflaton field $\varphi $ slowly
rolls along its potential and has an almost constant energy density which
drives an almost exponential expansion of the universal scale factor. The
comoving Hubble length (the horizon) decreases with time so that in this
process the quantum vacuum fluctuations of the inflaton field $\varphi $ on
the scales of interest to us leave the horizon (their scales become larger
than the horizon). The curvature perturbations they generate remain constant
during their subsequent super-horizon evolution, until these scales
eventually reenter the horizon long after the inflationary period has ended.
During the radiation-dominated era when these modes stay outside the
horizon, the metric perturbation $\Phi $ becomes a constant, which drives
the density perturbations of different matter species, and leads to the
observed CMB and matter power spectra after horizon re-entry.

There are no couplings between the scalar field and gravitational
or \text{\AE} -fields, so we can write down its dynamical
quantities for the scalar field $\varphi $ as
\begin{eqnarray}
\rho^{\varphi} &=& \frac{1}{2}\dot{\varphi}^{2} + V(\varphi),  \nonumber \\
p^{\varphi} &=& \frac{1}{2}\dot{\varphi}^{2} - V(\varphi),  \nonumber \\
q^{\varphi}_{a} &=& \dot{\varphi}\hat{\nabla}_{a}\varphi,  \nonumber \\
\Pi^{\varphi}_{ab} &=& 0.
\end{eqnarray}
Making the following harmonic expansion for $\tilde{\nabla}_{a}\varphi$
\begin{eqnarray}
\hat{\nabla}_{a}\varphi &=& \sum_{k}\frac{k}{a}\chi Q^{k}_{a},
\end{eqnarray}
it is easy to get
\begin{eqnarray}
\mathcal{X}^{\varphi} &=& \frac{1}{a^{2}}(\varphi^{\prime }\chi^{\prime }+
\varphi^{\prime 2}A + a^{2}V_{\varphi}\chi), \\
\mathcal{X}^{p, \varphi} &=& \frac{1}{a^{2}}(\varphi^{\prime }\chi^{\prime
}+ \varphi^{\prime 2}A - a^{2}V_{\varphi}\chi), \\
q^{\varphi} &=& \frac{1}{a^{2}}k\varphi^{\prime }\chi
\end{eqnarray}
where $V_{\varphi} \equiv \partial V(\varphi)/\partial\varphi$. Then
parallel to Eqs.~(\ref{eq:FIAdiabatic}, \ref{eq:FIAetherq}) we have,
following the standard procedure,
\begin{eqnarray}  \label{eq:FIAdiabaticq}
&& a^{2}\left(\mathcal{X}^{p, \varphi} - \frac{p^{\prime }}{\rho^{\prime }}%
\mathcal{X}^{\varphi}\right)  \nonumber \\
&=& \frac{4}{3}\left(1 + \frac{\Xi^{\prime }}{4\mathcal{H}\Xi}\right) \left[%
\varphi^{\prime }\tilde{\chi}^{\prime }- \varphi^{\prime 2}\Phi +
a^{2}V_{\varphi}\tilde{\chi} + 3\mathcal{H}\varphi^{\prime }\tilde{\chi}%
\right]\ \ \ \
\end{eqnarray}
and
\begin{eqnarray}  \label{eq:FIvarphiq}
\kappa a^{2}\tilde{q}^{\varphi} &=& \kappa[q^{\varphi} + (\rho^{\varphi} +
p^{\varphi})\sigma]  \nonumber \\
&=& \kappa k\varphi^{\prime }\tilde{\chi}
\end{eqnarray}
where we have defined the FI variable
\begin{eqnarray}
\tilde{\chi} &\equiv& \chi + \frac{\varphi^{\prime }}{k}\sigma.
\end{eqnarray}

Substituting Eqs.~(\ref{eq:FIAdiabatic}, \ref{eq:FIvarphiq}) into Eq.~(\ref%
{eq:PropagationPhi}), and using Eq.~(\ref{eq:Constraintphi2}) to eliminate
the term proportional to $\varphi^{\prime }\tilde{\chi}^{\prime
}-\varphi^{\prime }\Phi +a^{2}V_{\varphi}\tilde{\chi}+3\mathcal{H}%
\varphi^{\prime }\tilde{\chi}$, we arrive at the following equation
\begin{widetext}
\begin{eqnarray} \label{eq:PropagationPhi2}
&& \Phi'' + \left(2\mathcal{H}-\frac{\Xi'}{\Xi}\right)\Phi' +
\left(2\mathcal{H}'-\frac{\Xi'}{\Xi}\mathcal{H}\right)\Phi +
\frac{1}{1-\frac{1}{2}\alpha}\left(1+\frac{1}{2}c_{14}\right)k^{2}\Phi\nonumber\\
&=& \frac{1}{1-\frac{1}{2}\alpha}\left[\frac{1}{2}c_{14} -
\frac{1}{2}(c_{1}+c_{2}+c_{3}) - c_{13}\right]\frac{k}{a}\epsilon'
+ \frac{1}{1-\frac{1}{2}\alpha}
(c_{1}+c_{2}+c_{3})\frac{\Xi'}{2\Xi}\frac{k}{a}\epsilon -
\frac{1}{k}c_{13}\left(\frac{\epsilon'''}{a} -
\frac{\Xi'}{\Xi}\frac{\epsilon''}{a} -
\Xi\frac{\epsilon'}{a}\right).
\end{eqnarray}
\end{widetext}
Eqs.~(\ref{eq:Propagationepsilon}, \ref{eq:PropagationPhi2}) that form a
closed set of evolution equations for the coupled \text{\AE} -inflaton system;
they are a generalization of the results presented in \cite%
{AetherPerturbation}, and from them we can perform our analysis of the
observational effects.

Let us look first at the large-scale evolution of $\Phi$. In this limit $%
k\tau \ll 1$ so the last term on the left-hand side, and the first two terms
on the right hand side, of Eq.~(\ref{eq:PropagationPhi2}) can be dropped.
Notice that during the inflationary era $\mathcal{H}^{2}-\mathcal{H}^{\prime
}=\frac{1}{2}\kappa a^{2}(\rho+p)\simeq 0$ and Eq.~(\ref%
{eq:Propagationepsilon}) become
\begin{eqnarray}
(c_{14}-c_{13}\alpha)\epsilon^{\prime \prime }&=&
(\alpha+c_{14})k(a\Phi)^{\prime }.
\end{eqnarray}
Substituting this into Eq.~(\ref{eq:PropagationPhi2}) one could see that
\begin{eqnarray}
&& \frac{1}{k}c_{13}\left(\frac{\epsilon^{\prime \prime \prime }}{a} - \frac{%
\Xi^{\prime }}{\Xi}\frac{\epsilon^{\prime \prime }}{a} - \Xi\frac{%
\epsilon^{\prime }}{a}\right)  \nonumber \\
&\propto& \Phi^{\prime \prime }+ \left(2\mathcal{H}-\frac{\Xi^{\prime }}{\Xi}%
\right)\Phi^{\prime }+ \left(2\mathcal{H}^{\prime }-\frac{\Xi^{\prime }}{\Xi}
\mathcal{H}\right)\Phi  \nonumber
\end{eqnarray}
so that we finally have
\begin{eqnarray}  \label{eq:PropagationPhiLS}
\Phi^{\prime \prime }+ \left(2\mathcal{H}-\frac{\Xi^{\prime }}{\Xi}%
\right)\Phi^{\prime }+ \left(2\mathcal{H}^{\prime }-\frac{\Xi^{\prime }}{\Xi}%
\mathcal{H}\right)\Phi &=& 0
\end{eqnarray}
on super-horizon scales. Note that one should not simply set $%
\epsilon^{\prime \prime }=0$ in this limit as was done in Ref.~\cite%
{AetherPerturbation}, although that leads to the same result \cite{Comment2}%
. This equation is still valid \emph{outside} the inflationary
epoch within the parameter space given by
Eq.~(\ref{eq:ParameterSpace2}).

The solution to Eq.~(\ref{eq:PropagationPhiLS}) is given by
\begin{eqnarray}  \label{eq:LSPhi}
\Phi &=& D\left(1 - \frac{\mathcal{H}}{a^{2}}\int a^{2}d\tau\right)
\end{eqnarray}
where $D$ is a constant. Long after leaving the horizon, the potential $%
\Phi$ and the scalar perturbation $\tilde{\chi}$ have time to evolve, and
after inflation $\tilde{\chi}$ ceases to exist. However, as mentioned above,
during the radiation era on super-horizon scales $\Phi$ finally becomes a
constant which is related to $D$ by
\begin{eqnarray}
\Phi &=& \frac{2}{3}D.
\end{eqnarray}
Once $D$ is known we could set the initial conditions of $\Phi$ in the
radiation era for the subsequent evolutions. The quantity $D$ could be fixed
by matching to the inflaton perturbation $\tilde{\chi}$ at the time of
horizon exit ($k=aH=\mathcal{H}$) as follows: substituting Eqs.~(\ref%
{eq:FIAetherq}, \ref{eq:FIvarphiq}) into Eq.~(\ref{eq:Propagationphi2}) we
get
\begin{eqnarray}  \label{eq:PhiandInflaton}
k(a\Phi)^{\prime }&=& \frac{1}{1-\frac{1}{2}\alpha}\left[ \frac{1}{2}%
ka\kappa\varphi^{\prime }\tilde{\chi} - \frac{1}{2} (c_{1}+c_{2}+c_{3})k^{2}%
\epsilon\right]  \nonumber \\
&& - c_{13}\left[\epsilon^{\prime \prime }+ (\mathcal{H}^{\prime }-\mathcal{H%
}^{2})\epsilon\right].
\end{eqnarray}

Now we begin to confine ourselves within the parameter space Eq.~(\ref%
{eq:ParameterSpace}). In that case $c_{14}=-\alpha
=(c_{1}+c_{3})(c_{1}-c_{3})/c_{1}$ and the \text{\AE} -field EOM Eq.~(\ref%
{eq:Propagationepsilon}) becomes
\begin{eqnarray}
\alpha(1+c_{13})\left[\epsilon^{\prime \prime }+ (\mathcal{H}^{\prime }-
\mathcal{H}^{2})\epsilon\right] &=& (c_{1}+c_{2}+c_{3})k^{2}\epsilon.\ \ \
\end{eqnarray}
So, from these two equations we find that
\begin{eqnarray}
(a\Phi)^{\prime }&=& \frac{1}{1-\frac{1}{2}\alpha}\frac{1}{2}a
\kappa\varphi^{\prime }\tilde{\chi}.
\end{eqnarray}
On the other hand, from Eq.~(\ref{eq:LSPhi}) one can write
\begin{eqnarray}
(a\Phi)^{\prime }&=& \frac{1}{2}\frac{1}{1-\frac{1}{2}\alpha}\kappa
a\varphi^{\prime 2} \frac{1}{\mathcal{H}}D.
\end{eqnarray}
Obviously, matching these two expressions gives the value of $D$
\begin{eqnarray}
D &=& \frac{\mathcal{H}}{\varphi^{\prime }}\tilde{\chi}
\end{eqnarray}
where the three variables $\mathcal{H},\varphi^{\prime }$ and $\tilde{\chi}$
are all evaluated at the horizon exit. As a result, the initial power
spectrum for $\Phi$, $\mathcal{P}_{\Phi}=k^{3}\langle\Phi^{2}\rangle/2\pi
^{2}$ where $\langle\cdots\rangle$ means the ensemble average, is given by
\begin{eqnarray}  \label{eq:PPS}
\mathcal{P}_{\Phi} &=& \frac{4}{9} \frac{\mathcal{H}^{2}}{\varphi^{\prime 2}}%
\langle\tilde{\chi}^{2}\rangle  \nonumber \\
&=& \frac{4}{9}\frac{\mathcal{H}^{2}}{\varphi^{\prime 2}} \left(\frac{H}{2\pi%
}\right)^{2}
\end{eqnarray}
in which we have used the relation $\mathcal{P}_{\tilde{\chi}}=k^{3}\langle
\tilde{\chi}^{2}\rangle /2\pi ^{2}=(H/2\pi )^{2}$ \cite{InflationBook} (as
discussed in Ref.~\cite{AetherPerturbation}, this relation is not affected
by the \text{\AE} -field up to first order because the \text{\AE} -field is not coupled to
the inflaton) and the Hubble expansion rate $H=\dot{a}/a$ is also evaluated
at the time of horizon exit.

Note that the $\Phi $ power spectrum Eq.~(\ref{eq:PPS}) has exactly the same
form as in standard GR. To compare their magnitudes, let us write
\begin{eqnarray}
\frac{\mathcal{P}^{\text{\AE} }_{\Phi}}{\mathcal{P}^{\mathrm{GR}}_{\Phi}} &=& \frac{%
\left(H^{2}/\varphi^{\prime 2}\right)_{\text{\AE} }} {\left(H^{2}/\varphi^{\prime
2}\right)_{\mathrm{GR}}}  \nonumber
\end{eqnarray}
in which we have used the fact that at the time of horizon exit $\mathcal{H}%
=k$ should be the same in the two models. If we further assume the
same inflation potential, then \cite{Comment} because the
background expansion in our \AE ther model is indistinguishable
from that of GR, we have
\begin{eqnarray}
\frac{\mathcal{P}^{\text{\AE}
}_{\Phi}}{\mathcal{P}^{\mathrm{GR}}_{\Phi}} = 1,
\end{eqnarray}
that is, the power spectra in these two models are exactly the
same, in both the shape and the magnitude. It should be noticed
that if the background evolution, $H$, of the \AE-model during the
inflation is different from GR, then the evolutions of $\varphi$
will also be different because of the scalar field equation of
motion; as a result the slow-roll parameters at the horizon
crossing are generally different in these two models and so will
be the spectral indices.

What is the fate of the vector and tensor modes of the \text{\AE}
-field perturbation? As shown in \cite{AetherPerturbation}, the
vector mode is also sourceless unless exotic matter such as
cosmological defects exist. Hence, it should decay similarly all
the way up to the present and leave no traces in the CMB
observables. As for the tensor modes, from Eqs.~(\ref{eq:AErho} -
\ref{eq:AEpi}) one can see that the contribution of the \text{\AE}
-field to the tensor modes lies only in $\kappa\Pi^{\text{\AE}}$
and is independent of the model parameter $c_{4}$. Consequently,
the power spectrum of the tensor perturbations given in
\cite{AetherPerturbation}
\begin{eqnarray}
\mathcal{P}_{h} &=& \kappa(1+c_{13})^{1/2}\frac{H^{2}}{2\pi^{2}}
\end{eqnarray}
is still valid here (note that what appears in this formula is the
true bare gravitational constant). Following the argument above
again, we can show that
\begin{eqnarray}
\frac{\mathcal{P}^{\text{\AE} }_{h}}{\mathcal{P}^{\mathrm{GR}}_{h}} &=& \left(1+%
\frac{1}{2}c_{14}\right)(1+c_{13})^{1/2}.
\end{eqnarray}
According to Eq.~(\ref{eq:ParameterSpace2}), $c_{13}<0$ and $c_{14}<0$, so
the amplitude of the gravitational-wave spectrum predicted in our \text{\AE} -model
is smaller than that arising in standard GR. Although this may provide a
discrimination between the two models, the difficulty in observing the
tensor spectrum will be a great hurdle for the use of this discriminator: if
our Universe turns out to be described by the \text{\AE} -model, then this spectrum
is even more difficult to observe than in GR.

\subsection{The \text{\AE} -Effects on Late-time Cosmology}

We have seen in the above that in our \text{\AE} -model the primordial scalar-mode
power spectrum is exactly the same as assuming standard GR and using the
measured value $\kappa_{\star}$ in calculations. The next question is if
possible deviations from GR will emerge in the subsequent evolutions of the
perturbations. Interestingly, we find that the answer is yes. The reason is
that, although the \text{\AE} -perturbation is sourceless during the inflationary
era, in subsequent stages where the fluid anisotropic stresses are nonzero
it becomes sourced. To see the consequences explicitly, we now write the \text{\AE} %
-field EOM within the parameter space Eq.~(\ref{eq:ParameterSpace}) as
\begin{eqnarray}
\alpha\left[(1+c_{13})k(\tilde{\text{\ae} }^{\prime }+2\mathcal{H}\tilde{\text{\ae} }) +
\kappa\Pi^{f}a^{2}\right]^{\prime }&=& \frac{\alpha+2c_{13}}{3}k^{3}\tilde{%
\text{\ae} }. \nonumber
\end{eqnarray}

During the inflationary epoch $\Pi^{f} = 0$ and $\mathcal{H}^{\prime }-%
\mathcal{H}^{2} \simeq 0$, thus the solution to the above equation is
\begin{eqnarray}
\tilde{\text{\ae} }(\tau) &\simeq& \frac{A_{1}}{a}\sin\left(c_{s}k\tau\right) +
\frac{A_{2}}{a}\cos\left(c_{s}k\tau\right),
\end{eqnarray}
where we have defined $c^{2}_{s} \equiv \frac{c_{13}}{%
3(1+c_{13})(c_{1}-c_{3})}$ which is positive according to Eq.~(\ref%
{eq:ParameterSpace2}); $A_{1, 2}$ are integration constants. Meanwhile,
during this epoch the scale factor $a$ undergoes exponential growth, which
means that $\tilde{\text{\ae} }$ decays exponentially and its initial value is
washed out soon.

\begin{figure}[tbp]
\centering
\includegraphics[scale=0.9] {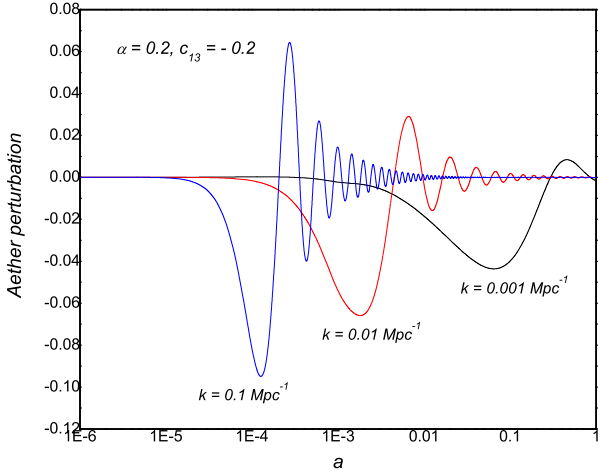}
\caption{(Color online) The evolution of the \text{\AE} -field perturbation $\tilde{%
\text{\ae} }$ versus the cosmic scale factor $a$. We have shown it
for 3 different scales $k = 0.1, 0.01, 0.001\ \mathrm{Mpc}^{-1}$
respectively, as indicated beside the curves. The model parameters
are $\protect\alpha = 0.2, c_{13} = -0.2$, which satisfy the
constraints Eq.~(\ref{eq:ParameterSpace2}).} \label{fig:Aether}
\end{figure}

In the radiation dominated epoch, $a \propto \tau$ and $\Pi^{f} =
\Pi^{\gamma} + \Pi^{\nu} \neq 0$ (here subscripts $_{\gamma},\ _{\nu}$
denotes photon and neutrino respectively). Now the homogeneous part of the
\text{\AE} -field EOM becomes
\begin{eqnarray}
\tilde{\text{\ae} }^{\prime \prime }+ \frac{2}{\tau}\tilde{\text{\ae} }^{\prime }- \frac{2%
}{\tau^{2}}\tilde{\text{\ae} } + c^{2}_{s}k^{2}\tilde{\text{\ae} } &=& 0
\end{eqnarray}
whose solution is
\begin{eqnarray}
\tilde{\text{\ae} }_{\mathrm{gen}}(\tau) &=& \frac{B_{1}}{\tau^{2}}\left[%
\cos(c_{s}k\tau)c_{s}k\tau - \sin(c_{s}k\tau)\right]  \nonumber \\
&& + \frac{B_{2}}{\tau^{2}}\left[\sin(c_{s}k\tau)c_{s}k\tau +
\cos(c_{s}k\tau)\right]
\end{eqnarray}
where $B_{1, 2}$ are integration constants and a subscript
$_{\mathrm{gen}}$ means the general solution. We can see that in
the limit $c_{s}k\tau \lesssim 1$ $
\tilde{\text{\ae}}_{\mathrm{gen}}$ decays as $\tilde{\text{\ae}
}_{\mathrm{gen}} \sim
1/\tau^{2}$ and when $c_{s}k\tau \gg 1$ it decreases as $\tilde{\text{\ae} }_{%
\mathrm{gen}} \sim B_{1}\cos(c_{s}k\tau)/\tau + B_{2}\sin(c_{s}k\tau)/\tau$.
Meanwhile there is a particular solution of $\tilde{\text{\ae} }$ which involves a
weighted integration of $(\kappa\Pi^{f}a^{2})^{\prime }$ over time. Deep
into the radiation-dominated epoch, the decaying general solutions of $%
\tilde{\text{\ae} }$ have become negligible and the growing
particular solution is still tiny, so we can reasonably take the
initial conditions of $\tilde{\text{\ae} } $ as $\tilde{\text{\ae}
}_{\mathrm{ini}} = \tilde{\text{\ae} }^{\prime }_{\mathrm{ini}} =
0 $ in our numerical calculations. We have checked that, for the
scales we are interested in, this choice of initial conditions is
robust and the numerical results are not sensitive to (not
dramatically) different initial conditions, and in
Fig.~\ref{fig:Aether} we have depicted the time evolution of
$\tilde{\text{\ae} }$ at different scales (or different $k$) for
the
model $\alpha = - c_{13} = 0.2$. It can be seen there that, at early times $%
\tilde{\text{\ae} }$ remains close to zero; later it grows as $(\kappa%
\Pi^{f}a^{2})^{\prime }$ deviates significantly from 0, and finally when $%
(\kappa\Pi^{f}a^{2})^{\prime }$ becomes tiny in the
matter-dominated era it undergoes a (oscillatory) decay again.

Now what kind of signatures will this behavior of
$\tilde{\text{\ae} }$ imprint on the perturbation evolution? In
order to answer this question, we find it useful to rewrite the
total density perturbation, pressure perturbation,
heat flux and anisotropic stress (with the contributions from the $\text{\AE} $%
-field included) as follows:
\begin{eqnarray}
\kappa\mathcal{X}a^{2} &=& \kappa\mathcal{X}_{\mathrm{tr}}a^{2} + \kappa%
\mathcal{X}_{\mathrm{ntr}}a^{2}, \\
\kappa\mathcal{X}^{p}a^{2} &=& \kappa\mathcal{X}^{p}_{\mathrm{tr}}a^{2} +
\kappa\mathcal{X}^{p}_{\mathrm{ntr}}a^{2}, \\
\kappa qa^{2} &=& \kappa q_{\mathrm{tr}}a^{2} + \kappa q_{\mathrm{ntr}}a^{2},
\\
\kappa\Pi a^{2} &=& \kappa\Pi_{\mathrm{tr}}a^{2} + \kappa\Pi_{\mathrm{ntr}%
}a^{2}
\end{eqnarray}
where the quantities with subscript $_{\mathrm{tr}}$
\begin{eqnarray}
\kappa\mathcal{X}_{\mathrm{tr}}a^{2} &=& \kappa_{\star}\mathcal{X}^{f}a^{2},
\\
\kappa\mathcal{X}^{p}_{\mathrm{tr}}a^{2} &=& \kappa_{\star}\mathcal{X}^{p,
f}a^{2}, \\
\kappa q_{\mathrm{tr}}a^{2} &=& \kappa_{\star}q^{f}a^{2}, \\
\kappa\Pi_{\mathrm{tr}}a^{2} &=& \kappa_{\star}\Pi^{f}a^{2}
\end{eqnarray}
are the tracking parts because they are exact rescalings of the
corresponding quantities for the fluid matter [we remind that
$\kappa_{\star} = \kappa/(1-\frac{1}{2}\alpha)$], while those with
subscript $_{\mathrm{ntr}} $ are the non-tracking parts expressed
as
\begin{eqnarray}
\kappa\mathcal{X}_{\mathrm{ntr}}a^{2} &=& \alpha\kappa_{\star}\Pi^{f}a^{2}
\nonumber \\
&& + \frac{\alpha}{1-\frac{1}{2}\alpha} (1+c_{13})k(\tilde{\text{\ae} }^{\prime }+ 2%
\mathcal{H}\tilde{\text{\ae} }), \\
\kappa\mathcal{X}^{p}_{\mathrm{ntr}}a^{2} &=& -\frac{1}{3}%
\alpha\kappa_{\star}\Pi^{f}a^{2}  \nonumber \\
&& - \frac{1}{3}\frac{\alpha}{1-\frac{1}{2}\alpha} (1+c_{13})k(\tilde{\text{\ae} }%
^{\prime }+ 2\mathcal{H}\tilde{\text{\ae} }), \\
\kappa q_{\mathrm{ntr}}a^{2} &=& -\frac{1}{3} \frac{\alpha+2c_{13}}{1-\frac{1%
}{2}\alpha}k^{2}\tilde{\text{\ae} }, \\
\kappa\Pi_{\mathrm{ntr}}a^{2} &=& c_{13}k(\tilde{\text{\ae} }^{\prime }+2\mathcal{H}%
\tilde{\text{\ae} }) -
\frac{1}{2}\alpha\kappa_{\star}\Pi^{f}a^{2};
\end{eqnarray}
only the measured value $\kappa_{\star}$ appears in the final
expressions. It can be easily checked that the tracking and
non-tracking parts satisfy the energy momentum conservation
equation separately by utilizing the \text{\AE} -field EOM. In
particular, the non-tracking part effectively represents a purely
perturbed contribution to the total energy
momentum tensor that has no background counterparts. We also note that $%
\kappa\mathcal{X}^{p}_{\mathrm{ntr}}a^{2} = -\frac{1}{3}\kappa\mathcal{X}_{%
\mathrm{ntr}}a^{2}$.

\begin{figure}[tbp]
\centering
\includegraphics[scale=0.9] {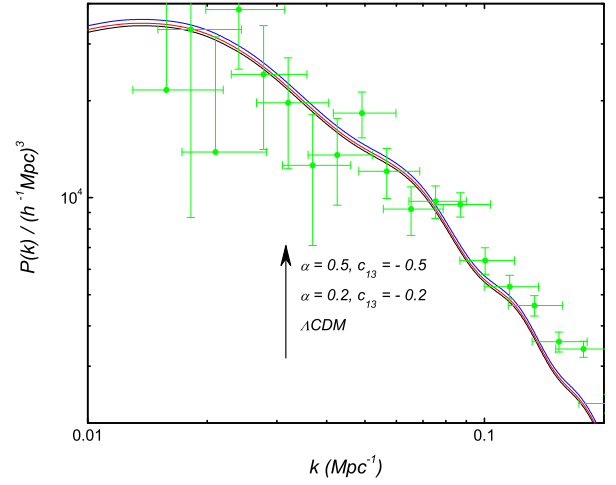}
\caption{(Color online) The matter power spectrum of the \text{\AE} -model
considered in this work. From top to bottom the curves correspond to $(%
\protect\alpha, c_{13}) = (0.5, -0.5), (0.2, -0.2)$ and $(0, 0)$
(which is just $\Lambda\mathrm{CDM}$) respectively. The other
parameters common for all curves are $\Omega_{b}h^{2} = 0.0223,
\Omega_{c}h^{2} = 0.1054$ and $h = 0.732$ where $\Omega_{b, c}$
are the fractional energy density of baryons and CDM, and $h
\equiv H_{0}/(100\ \mathrm{km/s/Mpc})$. The differences among
these spectra are small although their parameters are quite
different.} \label{fig:Pk}
\end{figure}

Taking the time derivative of Eq.~(\ref{eq:Constrainteta2}), using Eq.~(\ref%
{eq:Propagationeta2}) to eliminate the $\eta^{\prime }$ term and cancelling
the $\mathcal{X}^{\prime }$ appearing there with Eq.~(\ref%
{eq:EnergyConservation2}), one arrives at the following evolution equation
of $\mathcal{Z}$:
\begin{eqnarray}
k(\mathcal{Z}^{\prime }+\mathcal{HZ}) + \frac{1}{2}\kappa(\mathcal{X} + 3%
\mathcal{X}^{p})a^{2} &=& 0
\end{eqnarray}
or equally
\begin{eqnarray}
k(\mathcal{Z}^{\prime }+\mathcal{HZ}) + \frac{1}{2}\kappa_{\star}(\mathcal{X}%
^{f} + 3\mathcal{X}^{p, f})a^{2} &=& 0,
\end{eqnarray}
that is, the non-tracking part does not contribute to the
evolution of $\mathcal{Z}$. In this subsection we work in the CDM
frame where the 4-acceleration $A = 0$, and thus
$\Delta^{\prime}_{\mathbb{\mathrm{CDM}}} = -k\mathcal{Z}$
\cite{GR3+1} (where $\Delta \equiv \mathcal{X}/\rho$ is the
density contrast) which means the growth of matter perturbation is
not affected by the non-tracking part. However, since
$\kappa\mathcal{X}_{\mathrm{ntr}}a^{2}$, which is nonzero at early
times, appears in Eq.~(\ref{eq:Constrainteta2}), the values of
$\kappa_{\star}\mathcal{X}^{f}a^{2}$ and $\mathcal{Z}$ in this
equation are slightly modified that change the subsequent
evolution of $\mathcal{Z}$ a little. These can be seen in
Fig.~\ref{fig:Pk}, where we have displayed the matter power
spectrum of the \text{\AE} -model for different choices of
parameters. Obviously the spectrum depends weakly on the
parameters; furthermore, its shape is essentially the same as that
in $\Lambda\mathrm{CDM}$ model so that the parameters will
degenerate with the bias relating the matter power spectrum to the
observed spectrum of galaxies. In addition to that, there is also
the degeneracy with respect to the neutrino masses, since the
latter affects its anisotropic stress and the time at which they
become non-relativistic. Taken together this indicates that the
data for $P(k)$ will not place very stringent constraints on the
\text{\AE} -model.

\begin{figure}[tbp]
\centering
\includegraphics[scale=0.9] {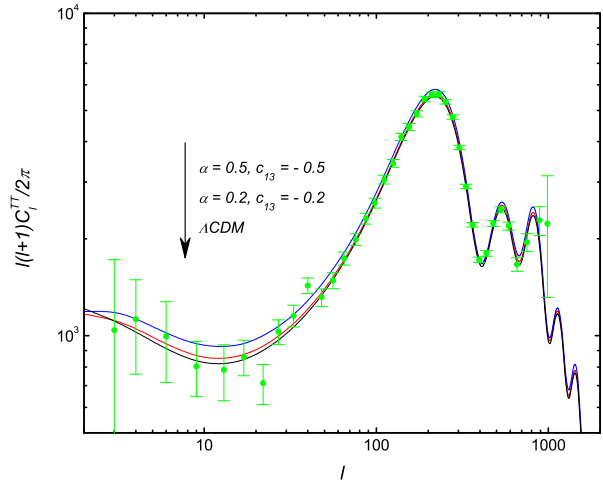}
\caption{(Color online) The CMB power spectrum of the \text{\AE} -model considered
in this work. At $l \sim 10$ from top to bottom the curves correspond to $(%
\protect\alpha, c_{13}) = (0.5, -0.5), (0.2, -0.2)$ and $(0, 0)$
(which is just $\Lambda\mathrm{CDM}$) respectively. The other
parameters common for all curves are $\Omega_{b}h^{2} = 0.0223,
\Omega_{c}h^{2} = 0.1054$ and $h = 0.732$ where $\Omega_{b, c}$
are the fractional energy density of baryons and CDM, and $h
\equiv H_{0}/(100\ \mathrm{km/s/Mpc})$.} \label{fig:CMB}
\end{figure}

From Eqs.~(\ref{eq:Constraintphi2} - \ref{eq:Propagationphi2}) one
can see that $\phi, \sigma^{\prime }, \phi^{\prime }$ are also
influenced by the non-tracking part, and because these variable
determine the CMB power spectrum, we could expect the latter to be
modified by the presence of the \text{\AE} -field as well. This is
correct, as shown Fig.~\ref{fig:CMB}, where we have plotted the
CMB spectra for different choices of the model. Obviously the CMB
power spectrum is changed by the \text{\AE} -field, though this
effect is not very strong and so it might be difficult to place
stringent limits on the parameter space of the theory.

There are two points that need to be noticed. Firstly, in this
work we are assuming that the recent cosmic acceleration is caused
by a cosmological constant, the anisotropic stress of which is
zero. If in contrast the dark energy has a nonzero anisotropic
stress (as in \cite{tomi,tomi2,jostein}) then the \text{\AE}
-field perturbation $\tilde{\text{\ae}}$ probably will not decay
at late times, and this might lead to further modifications of the
CMB power spectrum. Secondly, when $f(K)$ is not simply a linear
function of $K$, the \text{\AE} -field will be sourced by the
evolution of the gravitational potential as well, and its effects
will be more intricate and interesting. Such new features may
appear in more complicated models like TeVeS \cite{TeVeS,
TeVeSAether} and \text{\AE} -fields with non-canonical kinetic
terms \cite{NoncanonicalAether}.

In total, these results indicate that within the parameter space
Eq.~(\ref{eq:ParameterSpace}) the background cosmology of our
\text{\AE} -model is the same as in GR, while the CMB and matter
power spectra differ slightly from the predications of GR. In
principle, these features could be used to constrain the
parameters $\alpha$ and $c_{13}$ of the present model. This will
generally involve a full search of the parameter space using for
example the Markov Chain Monte Carlo method, which is beyond the
scope of this work. However, as we discussed above, the
constraints from the linear spectra may not be very stringent
anyway.

\section{Discussion and Conclusion}

\label{sect:Conclusion}

In this paper we have studied the cosmology of the
Einstein-\text{\AE} ther theory. After presenting the general
field equations for such theories in the CGI formalism, we
focussed on a specific class of models described in
Eq.~(\ref{eq:AELagrangian}), and confined ourselves to the
parameter space of models described by
Eq.~(\ref{eq:ParameterSpace2}) which pass the PPN and
$\mathrm{\breve{C}}$erenkov constraints. This parameter space is
known to have the same locally- and cosmologically-felt
gravitational constants ($\kappa _{0}=\kappa _{\infty }$, which
are different from the true bare $\kappa $) and this tracking
behavior indicates that we can consider its background expansion
just by using the measured value $\kappa _{\star }=\kappa _{0}$
and ignoring the presence of the \text{\AE} -field (since the only
effect of the \AE ther field is to track other matter species, the
model in Eq.~(\ref{eq:AELagrangian}) clearly cannot explain dark
energy. More general models, such as the $f(K)$ proposed in
\cite{NoncanonicalAether} and our Sec.~\ref{sect:Equations}, may
serve this purpose).

We find that it is a general feature that the tracking behavior
not only occurs at the background level but also at the linear
order in perturbation theory. For example, the quantity
$\mathcal{X}^{p,\text{\AE}}-\frac{p^{\prime\text{\AE}}}{\rho^{\prime
\text{\AE} }}\mathcal{X}^{\text{\AE} }$ tracks that of the other
matter species on super-horizon scales. This indicates that,
whatever type of perturbation is generated during inflation, the
presence of \text{\AE} -field will not alter it. In particular, in
the single-field inflation model we consider, no isocurvature
perturbation is produced. This is an important characteristic, and
it would be interesting to see whether similar behavior occurs for
general, higher-order choices of $f(K)$.

We generalized the analysis of primordial power spectra for the
\text{\AE} -theory \cite{AetherPerturbation} to our model. For the
parameter ranges which satisfy local gravity bounds, we find that
the evolution of the large-scale gravitational potential, $\Phi$,
is unmodified as compared with that in GR, and show that the
primordial power spectrum of $\Phi$ also has the same form as in
GR [c.f.~Eq.~(\ref{eq:PPS})]. If we assume that the bare
gravitational constants in GR and in the \text{\AE} -model are the
same, then the magnitudes of the spectra are different in these
two models. However, contrary to the discussion in
\cite{AetherPerturbation}, we argue that we do not know the true
bare $\kappa$, but only know the measured value, $\kappa_{\star} =
\kappa_{0}$. In both GR and our \text{\AE} -model, $\kappa_{0}$ is
equal to the cosmological value $\kappa_{\infty},$ while in the
latter $\kappa \neq \kappa_{\infty}$. As a result, we show that
with the same inflationary potential the primordial power spectra
of $\Phi$ in these two models should have the same shape and the
same magnitude. Meanwhile, we also find that the power spectrum of
tensor perturbations in our model is smaller in magnitude than is
predicted in GR, and so currently we cannot use it to distinguish
between GR and the \text{\AE} -model.

For the late-time evolution of the perturbations, it is shown
that, the \text{\AE} -field perturbation is driven by the
evolution of the anisotropic photon and neutrino stresses -- the
$[\kappa_{\star }(\Pi^{\gamma}+\Pi^{\nu})a^{2}]^{\prime}$ term in
its propagation equation -- although it is sourceless during
inflation. Therefore, it decays away exponentially during the
inflationary era, grows again in the radiation-dominated epoch
when $[\kappa_{\star}(\Pi^{\gamma}+\Pi^{\nu})a^{2}]^{\prime}$ is
significant, and finally diminishes again, oscillating when
$[\kappa_{\star}(\Pi^{\gamma}+\Pi^{\nu})a^{2}]^{\prime}$
eventually becomes negligible. As a result, the CMB and matter
power spectra are modified by the existence of the \text{\AE}
-field. We also remark that, depending on the nature of the dark
energy, it is possible that the \text{\AE} -field perturbation
will have non-trivial dynamics driven by the
$(\kappa_{\star}\Pi^{\mathrm{DE}}a^{2})^{\prime}$ term, where
$\Pi^{\mathrm{DE}}$ is the possible dark-energy anisotropic
stress \cite{tomi}.

We also note that there recently appeared a later paper
\cite{ZFSLSS} which investigated in details the structure
formation of the model proposed in \cite{NoncanonicalAether}. In
their Sec.~V the authors considered a simple power-law model $f(K)
= \gamma(-K)^{n}$ in which $n = 1$ corresponds to our Lagrangian
Eq.~(\ref{eq:AELagrangian}) (except for the $c_{4}$ term). They
obtained criteria for there to be a growing mode in the evolution
of the (scalar-mode) \AE-field perturbation. However their
criteria cannot be applied directly to the model we consider here.
To see why, note that in Eq.~(32) of \cite{ZFSLSS} the quantities
$\Psi, \Phi$ also contain $V$ (their $V$ is equivalently our
$\tilde{\ae}$). However, by their argument the $V$ terms in $\Psi,
\Phi$ are suppressed by a small quantity $F_{K}$ (which in
\emph{our} notation is just $F$) and are ultimately neglected from
the \AE-field equation of motion, their Eq.~(43). In our model,
there is no argument that $F \ll 1$ (its magnitude actually
\emph{is} 1) and so we can no longer neglect the $\tilde{\ae}$
terms appearing in $\Phi, \Psi$. When these are taken into account
we obtain a different equation of motion, \emph{i.e.}, the
$b_{i}$s in Eq.~(43) of \cite{ZFSLSS} are different in our work,
and consequently the criteria for the existence of growing modes
are different as well. In addition, are the facts that we have a
$c_{4}$ term in our Lagrangian and use a different parameter
space, Eqs.~(\ref{eq:ParameterSpace}, \ref{eq:ParameterSpace2}),
which also contribute to the differences between our results and
those in \cite{ZFSLSS}. As an aside, we stress that the key
relation in our work, $c_{14} = -\alpha$, is a consequence of
Eq.~(\ref{eq:ParameterSpace}) and has nothing to do with whether
or not the \AE-gravity waves propagate superluminally. In fact, we
could do our calculation dropping the constraint that these waves
propagate superluminally, and in this case the perturbation
dynamics might place some constraints on the parameter space; for
example, in some portion of the parameter space the growth of
$\tilde{\ae}$ may become unstable. In our work we choose the
parameters leading to superluminal gravity waves simply to avoid
constraints from $\breve{\mathrm{C}}$erenkov radiation. There are
still some debates about this kind of choice, see however
\cite{AetherWF} for a different and conservative point of view.

There are also other general differences between the EA model we
consider here and the GEA model of \cite{ZFSLSS}. First, for the
static and weak field limit, the $c_{4}$ term and the choice of
parameter space Eq.~(\ref{eq:ParameterSpace}) are crucial for the
EA model to evade PPN tests \cite{AetherWF}; in \cite{ZFSLSS}
there is no $c_{4}$ term, but the nonlinearity in the \AE ther
field Lagrangian guarantees that the local gravitational tests are
not a problem because at high densities the modification is
negligible. Second, as we have seen above, our conclusion that the
primordial power spectrum of density perturbation is unmodified
compared with GR relies on the fact $c_{14} = -\alpha$ so that the
background expansion during inflation in our model is the same as
in GR (again the $c_{4}$ term is crucial here). If the $c_{4}$
term is not included, the local and cosmological gravitational
constants are different, meaning that the background evolution
during inflation is different from GR; as a result the background
quantities at horizon crossing, which determine the observables in
inflationary models through the slow-roll parameters, may be
different from those in GR -- this means that the EA model will
predict different shape of primordial density spectrum other than
GR \cite{Limcommunication}. For the GEA model considered in
\cite{ZFSLSS}, again the nonlinear $f(K)$ makes the modifications
to GR during early times small enough to be neglected, and as such
the primordial power spectrum is also the same as in GR. Note
however that in the latter case \cite{ZFSLSS} the late time
evolution of vector field perturbation is also modified in a
complex way, making the corresponding cosmological behaviors
different from ours.

So, in conclusion, although the background expansion of our
\text{\AE} -model is exactly the same as that in GR, the
cosmological data on CMB and matter power spectra can be used to
constrain the \text{\AE} -model. However, the constraint is not
expected to be very stringent because the linear perturbation
spectra depend weakly on the model parameters. Other
considerations of the behaviour in strong gravitational fields,
such as those studies of the compact stars or black holes
\cite{AetherCompactStars, AetherBHs} would enable better
constraints on the present model to be obtained. Alao, as general
interests for cosmology, studies in more complicated \AE-field
Lagrangian as those performed in \cite{ZFSLSS, Zhao2007b} need to
be explored thoroughly.

\begin{acknowledgments}
We thank Ted Jacobson, Eugene Lim, Constantinos Skordis, Xiaoting
Wang, HongSheng Zhao, Tom Zlosnik and the referee for helpful
discussions. The numerical calculation in this work uses a
modified version of the public CAMB code \cite{CAMB}. BL is
supported by the Overseas Research Studentship, Cambridge Overseas
Trust and the DAMTP at University of Cambridge. DFM acknowledges
the Humboldt Foundation.
\end{acknowledgments}


\begin{thebibliography}
\bibitem[Comment(2007)]{Comment} We want to elaborate this point as it will
also be used below. Suppose we have a universe described by either standard
GR or the \text{\AE} -theory Eq.~(\ref{eq:AELagrangian}) and we have a measured
gravitational constant $\kappa _{\star }$. If we assume GR, then we have $%
\kappa _{0}^{\mathrm{GR}}=\kappa _{\star }=\kappa ^{\mathrm{GR}}=\kappa
_{\infty }^{\mathrm{GR}}$, while otherwise we have $\kappa _{0}^{\text{\AE} %
}=\kappa _{\star }=\frac{1}{1+\frac{1}{2}c_{14}}\kappa ^{\text{\AE} }=\frac{1-\frac{%
1}{2}\alpha }{1+\frac{1}{2}c_{14}}\kappa _{\infty }^{\text{\AE} }$, so that
whenever we need $\kappa _{\infty }$ we could relate it to the
textbook-value $\kappa _{\star }$ by $\kappa _{\infty }=\kappa _{\star }$
(in the GR case), or $\kappa _{\infty }=\frac{1+\frac{1}{2}c_{14}}{1-\frac{1%
}{2}\alpha }\kappa _{\star }$ (in the \text{\AE} -case). Obviously, when $%
c_{14}=-\alpha $, $\kappa _{\infty }^{\text{\AE} }=\kappa _{\infty }^{\mathrm{GR}},$
 the background cosmologies are the same in GR and our \text{\AE} -model.
\bibitem[Comment2(2007)]{Comment2}

With general choices of parameters $c_{1,2,3,4}$, Eq.~(\ref%
{eq:PropagationPhiLS}) is only valid if $\mathcal{H}^{\prime }-\mathcal{H}%
^{2}\simeq 0$. Thus, in the radiation-dominated era it cannot apply even for
superhorizon scales, and its solution Eq.~(\ref{eq:LSPhi}) should be
modified as well.
\end{thebibliography}
\end{document}